\documentclass[10pt, a4paper, twocolumn, twoside]{article} 

\usepackage[english]{babel} 

\usepackage{microtype} 

\usepackage{amsmath,amsfonts,amsthm} 

\usepackage[svgnames]{xcolor} 

\usepackage[small, labelfont=bf, up]{caption} 

\usepackage{booktabs} 

\usepackage{lastpage} 

\usepackage{graphicx} 

\usepackage{enumitem} 
\setlist{noitemsep} 

\usepackage{sectsty} 
\allsectionsfont{\usefont{OT1}{phv}{b}{n}} 

\usepackage{geometry} 

\usepackage[T1]{fontenc} 
\usepackage[utf8]{inputenc} 

\usepackage{XCharter} 

\usepackage{hyperref}

\usepackage{journals}

\usepackage{fancyhdr} 
\newcommand{\shorttitle}[1]{\fancyhead[CE]{\textsl{#1}}}
\newcommand{\shortauthors}[1]{\fancyhead[CO]{\textsl{#1}}}

\fancypagestyle{firstpage}{ 
	\fancyhf{}
        \lhead{\vspace*{0.0em} \textit{\footnotesize 21$^{\rm st}$ European Workshop on White
            Dwarfs \\ Castanheira, Campos, Montgomery, eds. \\[-0.2em]
          July 23--27, 2018, Austin, Texas, USA}}
}

\date{}

\newcommand{\authorstyle}[1]{{\large\usefont{OT1}{phv}{b}{n}\color{DarkRed}#1}} 

\newcommand{\institution}[1]{{\footnotesize\usefont{OT1}{phv}{m}{sl}\color{Black}#1}} 

\usepackage{titling} 

\newcommand{\HorRule}{\color{DarkGoldenrod}\rule{\linewidth}{1pt}} 

\pretitle{
	\vspace{-30pt} 
	\HorRule\vspace{10pt} 
	\fontsize{16}{12}\usefont{OT1}{phv}{b}{n}\selectfont 
	\color{DarkRed} 
}

\posttitle{\par\vskip 10pt} 

\preauthor{} 

\postauthor{ 
	\vspace{0pt} 
	{\par\HorRule} 
	\vspace{-10pt} 
}

\newcommand{\newabstract}[1]{
    {\section*{Abstract}
    \bfseries #1}
  }

\usepackage[authoryear]{natbib}
\title{Light variability of white dwarfs and subdwarfs due to
surface abundance spots}

\shorttitle{Light variability of white dwarfs and subdwarfs due to
surface abundance spots}

\shortauthors{Krti{\v c}ka, Prv\'ak, Krti\v{c}kov\'a, Mikul\'a\v sek and
A.~Kawka}

\author{
        \authorstyle{J.~Krti\v{c}ka$^1$, M.~Prv\'ak$^1$, I.~Krti\v{c}kov\'a$^1$,
        Z.~Mikul\'a\v sek$^1$, A.~Kawka$^2$}
	\newline\newline 
	$^1$\institution{Department of Theoretical Physics and Astrophysics,
        Masaryk University, Brno, Czech Republic}\\
	$^2$\institution{International Centre for Radio Astronomy Research,
        Curtin University, Perth, Australia}
      }

\begin{document}

\maketitle 

\thispagestyle{firstpage} 


\newabstract{ Classical main-sequence chemically peculiar stars show light
variability that originates in surface abundance spots. In the spots, the flux
redistribution due to line (bound-bound) and bound-free transitions is modulated
by stellar rotation and leads to light variability. White dwarfs and hot
subdwarfs may also have surface abundance spots either owing to the elemental
diffusion or as a result of accretion of debris. We model the light variability
of typical white dwarfs and hot subdwarfs that results from putative surface
abundance spots. We show that the spots with radiatively supported iron
overabundance may cause observable light variability of hot white dwarfs and
subdwarfs. Accretion of debris material may lead to detectable light variability
in warm white dwarfs. We apply our model to the helium star HD 144941 and
conclude that the spot model is able to explain most of observed light
variations of this star. }


\section{Motivation: Main-sequence Chemically Peculiar Stars}

The current paradigm predicts that the photometric variability of main-sequence
chemically peculiar stars is caused by the flux redistribution (blanketing) in
surface abundance patches modulated by stellar rotation. The distribution of
chemical elements over the surface of chemically peculiar stars is typically not
uniform. The elements concentrate in vast persistent patches, partially governed
by the magnetic field. The overabundance of elements leads to the flux
redistribution from the far-UV to near-UV and visible domains due to the
bound-bound (lines, mainly iron or chromium) and bound-free (ionization, mainly
silicon and helium) transitions. Consequently, as a result of their rotation, CP
stars show magnetic, spectral, and photometric variability
\citep[e.g.,][]{mycuvir,myphidra}.

The abundance anomalies in main-sequence chemically peculiar stars are caused by
the radiative diffusion in quiet stellar atmospheres. The same processes can
also be important in hot subdwarfs or in hot white dwarfs. Moreover, surface
abundance spots in white dwarfs may be connected with the accretion of debris
material. All these processes can lead to the photometric variability, which was
detected in evolved compact stars \citep[e.g.,][]{dupy,kilic}.


\section{Model of Variability}

\begin{figure}
  \centerline{\includegraphics[angle=0,width=\columnwidth]{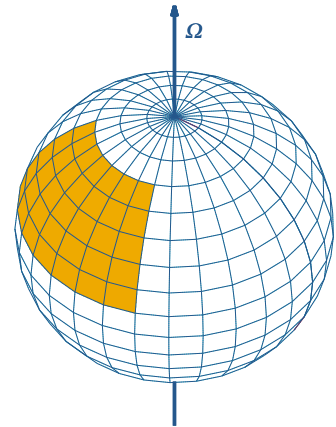}}
  \caption{Adopted spot model.}
  \label{povrch}
\end{figure}

We assume the existence of a spot with an overabundance of heavy elements on
the stellar surface (Fig.~\ref{povrch}). The spot has latitude
$10^\circ\leq\vartheta\leq70^\circ$ and longitude
$180^\circ<\varphi<270^\circ$. We model the influence of the spot on the
emergent radiation using the TLUSTY code \citep{bstar2006}.

\begin{figure}
  \centerline{\includegraphics[angle=0,width=\columnwidth]{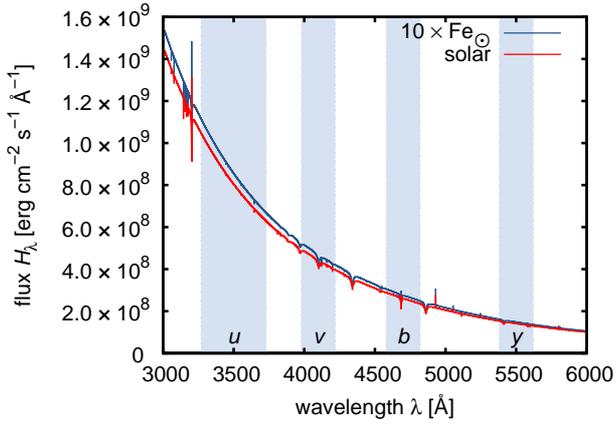}}
  \caption{Emergent flux from model atmospheres with two different chemical 
  compositions.}
  \label{magtoky}
\end{figure}

Higher abundance of heavy elements causes stronger blocking of radiation in the
UV. This leads to a redistribution of the radiative flux. Part of the
redistributed flux appears in the optical region and causes brightening of the
star, for example, in Str\"omgren colours as shown in Fig.~\ref{magtoky} for a
hot white dwarf with $T_\text{eff}=100\,231\,$K and with overabundance of iron. The
rotationally modulated brightening can be detected via optical photometry.

\section{Hot White Dwarfs: Diffusion}

\begin{figure}
  \centerline{\includegraphics[angle=0,width=\columnwidth]{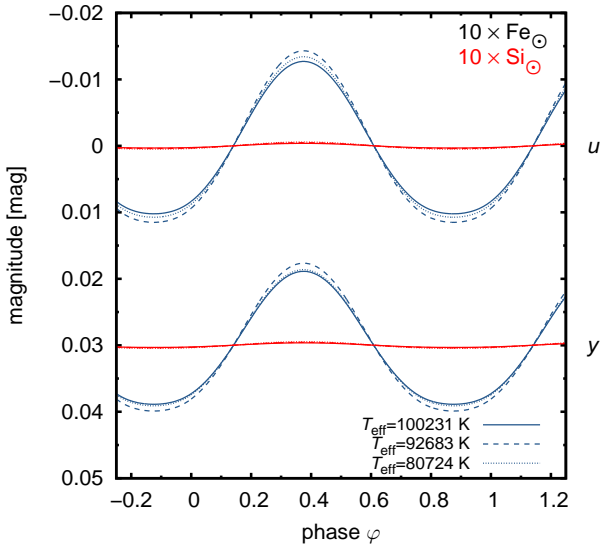}}
  \caption{Calculated light curves of hot white dwarfs on the top of white dwarf cooling track.}
  \label{abtvit}
\end{figure}

\begin{figure}
  \centerline{\includegraphics[angle=0,width=\columnwidth]{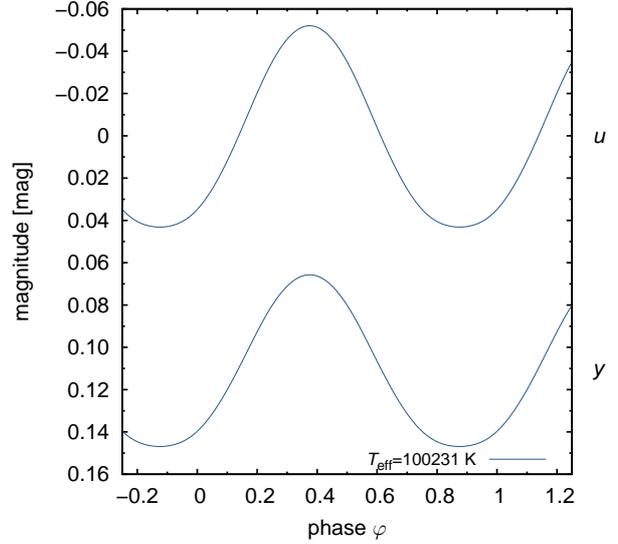}}
  \caption{Calculated light curves of hot white dwarf on the top of white dwarf
  cooling track with pure hydrogen atmosphere and solar metallicity spot.}
  \label{t100231g699y0851HAbtvit}
\end{figure}

For the calculation of light curves in Fig.~\ref{abtvit} we assumed spots with
an abundance of iron and silicon $10\times$ higher than the solar value. While
the silicon spot does not cause any significant light variability, the iron
spot leads to variability with amplitude of the order of 0.01~mag. Such
variations could be easily detected from observations. The simulated light
curves are nearly the same in all colours of the Str\"omgren photometric
system. This happens because at such high temperatures the optical region lies
in the Rayleigh-Jeans part of the flux distribution. Even stronger variability
appears with solar metallicity spot on purely hydrogen surface mostly due to
helium (Fig.~\ref{t100231g699y0851HAbtvit}).


\section{Warm White Dwarfs: Accretion}

\begin{figure}
  \centerline{\includegraphics[angle=0,width=\columnwidth]{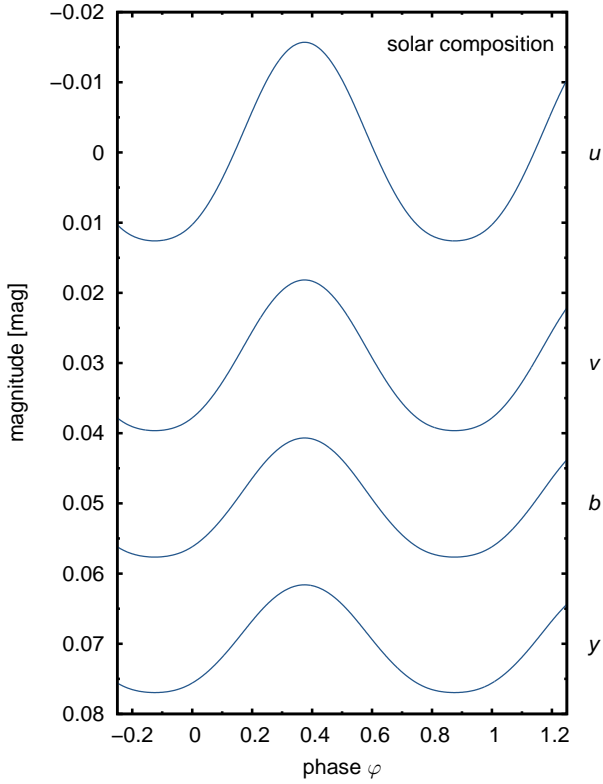}}
  \caption{Simulated light curves of a warm white dwarf ($T_\text{eff}=20\,$kK).}
  \label{habtvit}
\end{figure}

The calculations of the light curves of a warm white dwarf assume a spot with a
solar abundance of heavy elements while the rest of the surface is composed of
pure hydrogen. Such a spot causes light variability with an amplitude of a few
0.01~mag, which decreases with increasing wavelength. These variations could be
detected even from ground-based observations.

\section{Hot Subdwarfs: Diffusion}

\begin{figure}
  \centerline{\includegraphics[angle=0,width=\columnwidth]{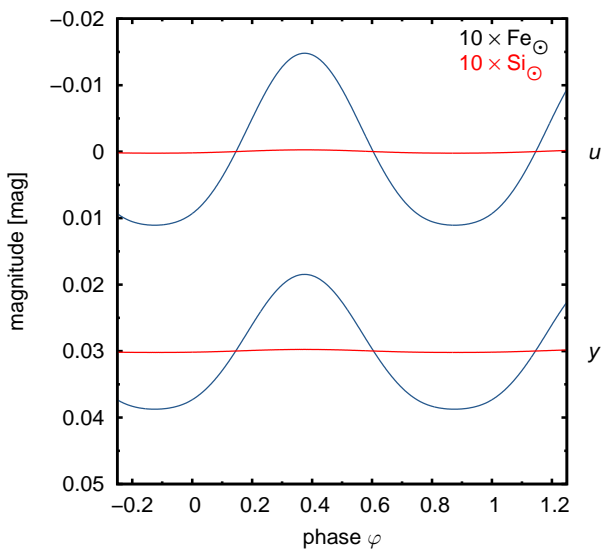}}
  \caption{Simulated light curves of a hot subdwarf with $T_\text{eff}=44\,$kK.}
  \label{hd49798}
\end{figure}

For the calculation of light curves of a hot subdwarf we assumed spots with an
abundance of iron and silicon $10\times$ higher than the solar value. The rest
of the stellar surface has solar chemical composition. The silicon spot does
not cause any significant light variability, whereas the iron spot causes light
variability with an amplitude of the order of 0.01~mag. The variations due to
the iron spot could be easily detected.

\section{HD~144941: a test against reality}

\begin{figure}
  \centerline{\includegraphics[angle=0,width=\columnwidth]{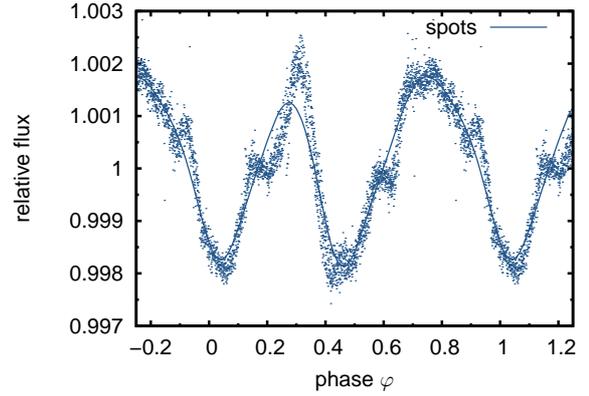}}
  \caption{Observed and simulated light curve of HD~144941.}
  \label{hd144941}
\end{figure}

\citet{jefram}
attributed the light variability of helium star HD 144941 to surface spots. We
attempted to find a surface spot distribution that provides the best fit of the
observed K2 light curve. We assumed maximum spot brightness variations of
$7.5\,\%$. This can explain the main observed trends, while the remainder could
be possibly caused by the light absorption in circumstellar clouds
\citep{labor}.

\section{Conclusions}

White dwarfs and hot subdwarfs may show rotationally modulated light variability
due to flux redistribution in surface abundance spots. The spots may originate
as a result of radiative diffusion in hot white dwarfs and subdwarfs or due to
accretion in warm white dwarfs.

\section*{Acknowledgement} We thank Dr.~Stephane Vennes for useful
discussions. This research was supported by grant GA\,\v{C}R 16-01116S.


\bibliography{krticka2_papers}

\end{document}